\documentclass[journal]{ASD}
\renewcommand\figurename{Figure}
\renewcommand\refname{References}
\renewcommand\tablename{Table}
\IEEEoverridecommandlockouts

\IEEEoverridecommandlockouts

\usepackage[utf8]{inputenc} % Kullanılan encodinge göre utf8 yerine latin5 de yazılabilir.
\usepackage[T1]{fontenc}

\usepackage{cite}
\ifCLASSINFOpdf
  \usepackage[pdftex]{graphicx}
\else
\fi
\usepackage[cmex10]{amsmath}
\usepackage{multirow}
\usepackage{array}
\usepackage[lofdepth,lotdepth]{subfig}

\begin{document}

%% Bu kısım IEEE'de yayımlanıyorsa aktif ediliyor.
%\IEEEpubid{\makebox[\columnwidth]{ 978-1-5090-2386-8/16/\$31.00 ©2016 IEEE\hfill}
%\hspace{\columnsep}\makebox[\columnwidth]{}}

\title{Effect of Different Batch Size Parameters on Predicting of COVID19 Cases}
%Farklı Parti Boyutu Parametrelerinin COVID19 Vakalarının Tahmini Üzerindeki Etkisi}

\author{\IEEEauthorblockN{Ali Narin$^{1}$, Ziynet Pamuk$^{2}$}\\
\IEEEauthorblockA{$^{1}$Electrical and Electronics Enginering, Zonguldak Bulent Ecevit University, Zonguldak, Turkey\\
$^{2}$Biomedical Enginering, Zonguldak Bulent Ecevit University, Zonguldak, Turkey\\}
{\{alinarin,ziynet.pamuk\}}@beun.edu.tr}
% journal papers do not typically use \thanks and this command
% is locked out in journal mode. If really needed, such as for


\maketitle

\begin{abstract}
The new coronavirus 2019, also known as COVID19, is a very serious epidemic that has killed thousands or even millions of people since December 2019. It was defined as a pandemic by the world health organization in March 2020.  It is stated that this virus is usually transmitted by droplets caused by sneezing or coughing, or by touching infected surfaces. The presence of the virus is detected by real-time reverse transcriptase polymerase chain reaction (rRT-PCR) tests with the help of a swab taken from the nose or throat. In addition, X-ray and CT imaging methods are also used to support this method. Since it is known that the accuracy sensitivity in rRT-PCR test is low, auxiliary diagnostic methods have a very important place. Computer-aided diagnosis and detection systems are developed especially with the help of X-ray and CT images. Studies on the detection of COVID19 in the literature are increasing day by day. In this study, the effect of different batch size (BH=3, 10, 20, 30, 40, and 50) parameter values on their performance in detecting COVID19 and other classes was investigated using data belonging to 4 different (Viral Pneumonia, COVID19, Normal, Bacterial Pneumonia) classes. The study was carried out using a pre-trained ResNet50 convolutional neural network. According to the obtained results, they performed closely on the training and test data. However, it was observed that the steady state in the test data was delayed as the batch size value increased. The highest COVID19 detection was 95.17\% for BH = 3, while the overall accuracy value was 97.97\% with BH = 20. According to the findings, it can be said that the batch size value does not affect the overall performance significantly, but the increase in the batch size value delays obtaining stable results.
%\boldmath
\end{abstract}

\IEEEpeerreviewmaketitle

\IEEEpubidadjcol

\section{INTRODUCTION}

Epidemic and seasonal flu infections cause severe pneumonia and high mortality rates \cite{Zou2020}. Pneumonia is an inflammation of the lung tissue \cite{Sirazitdinov2019,Saul2020}. Thus, human breathing becomes painful and oxygen intake decreases. Pneumonia occurs due to viruses, bacteria, and fungi \cite{Verma2020}.
Nowadays the whole world is struggling with the COVID19 epidemic. Deaths from pneumonia developing due to the SARS-COV-2 virus are increasing day by day. The vast majority (99\%) of people suffering from COVID19 epidemic disease have mild symptoms \cite{World2020}. Although the remaining 1\% seems small, it actually refers to millions of people. These people suffer from pneumonia caused by SARS-COV-2. Pandemic captured elderly, obese, pregnant women or people with chronic diseases should be admitted in the intensive care unit of the hospital. Mortality rates of these patients are high \cite{Zou2020}. According to the World Health Organization's 2019 report, pneumonia ranks first among infection-related deaths worldwide \cite{WHO2020}.
The most important method of diagnosis of pneumonia worldwide is chest X-ray \cite{Jaiswal2019}. Doctors make a diagnosis by looking at a chest X-ray. However, doctors deal with many patients and work long hours in disasters such as epidemics. At the same time, healthcare workers have a very high risk of getting sick. For such reasons, engineers working with artificial intelligence work on decision support systems to facilitate the work of doctors.
In recent years, Deep Learning (DL) technique has been able to diagnose diseases (e.g. lung disease, metastasis detection for breast cancer, skin lesion classification, diabetic retinopathy, attention deficit hyperactivity disorder) just like specialist physicians \cite{Ghoshal2020}. DL is one of the most successful artificial intelligence techniques and is an effective tool to help radiologists analyze \cite{Zhang2020}. There is interest in using convolutional neural networks (CNNs) to analyze medical imaging to provide computer-aided diagnosis (CAD) \cite{Zech2018}. CNNs are one of the powerful deep learning architectures. CNN is intuitively widely used in many practical applications such as pattern recognition and image classification \cite{Hemdan2020}. 
CAD systems based on deep learning and medical imaging are becoming more and more research centers \cite{Liang2020}. Verma and others suggest a CNN model prepared without any preparation to group and identify the occurrence of pneumonia disease from a particular chest X-ray image test product line \cite{Verma2020}. Gusztáv Gaál and colleagues provide a new approach to deep learning to diagnose pneumonia, lung segmentation, a basic but challenging task. In this system, state-of-the-art fully CNNs are used together with a competing critic model \cite{Gaal2020}. In the study of Ilyas Sirazitdinov, he explores opportunities to apply machine learning solutions for the automatic detection and localization of pneumonia in chest X-ray images. RetinaNet and Mask R-CNN are recommended for the detection and localization of pneumonia \cite{Sirazitdinov2019}. Wesley O’Quinn and his friends have detected pneumonia using chest X-rays and AlexNet transfer network in deep learning \cite{Quinn2019}. Stephan and others proposed a CNN model trained from scratch to classify the presence of pneumonia from chest X-ray images \cite{Stephan2019}. Sreeja and others used the neural network classification model in Keras to classify Normal and Pneumonia using lung X-ray data \cite{Sreeja2019}. Longjiang and others evaluate the effectiveness of the deep learning model in segmenting the lung and thorax regions on pediatric chest X-rays (CXR) \cite{Longjiang2019}. For the diagnosis of pneumonia, M.Toğaçar and others use existing CNNs, which are AlexNet, VGG16 and VGG19 \cite{Togacar2019}. Apostopolus and others performed a common pneumonia, COVID19-induced pneumonia, and an evolutionary neural network for healthy differentiation for automatic detection of COVID19. In particular, the procedure called transfer learning has been adopted. With transfer learning, the detection of various abnormalities in small medical image datasets is an achievable goal, often with remarkable results \cite{Apostopolus2020}. Based on chest X-ray images, Zhang and others aim to develop a deep learning-based model that can detect COVID19 with high sensitivity, providing fast and reliable scanning \cite{Zhang2020}. Ghoshal and Tucker used the dropweights-based Bayesian CNNs model using chest X-ray images for diagnosis of COVID19 \cite{Ghoshal2020}. Hemdan and others used VGG19 and DenseNet models to diagnose COVID19 from X-ray images \cite{Hemdan2020}. Apostopolus and others performed automatic detection from X-ray images using CNNs with Transfer Learning \cite{ApostopolusTzani2020}. In another study, Apostopolus used the latest technology CNN called MobileNet and used 7 classes \cite{Apostopolus2020}.  Sethy and Behera proposed a methodology based on Deep Learning to detect a patient infected with coronavirus using X-ray images. In addition to the ResNet50 method, they also worked on the SVM model \cite{Sethy2020}. 

In our first study in the diagnosis of pneumonia developed due to COVID19, we classified the chest x-ray images using CNN-based ResNet50, InceptionV3 and Inception-ResNetV2 models. As a result of this study, ResNet50 model achieved the highest performance with 98\% classification performance \cite{Narin2020}. Based on the results obtained from this study, the effects of the different bacthsize hyper-parameter on the performance of the COVID19 cases were examined. For this, pre-trained ResNet50 model and X-ray images in 4 classes (COVID19, Normal, Viral Pneumonia, Bacterial Pneumonia) were used. The flow chart of the proposed study is given in Figure 1.

\begin{figure}[htbp]
\centering
\includegraphics[width=0.95\columnwidth]{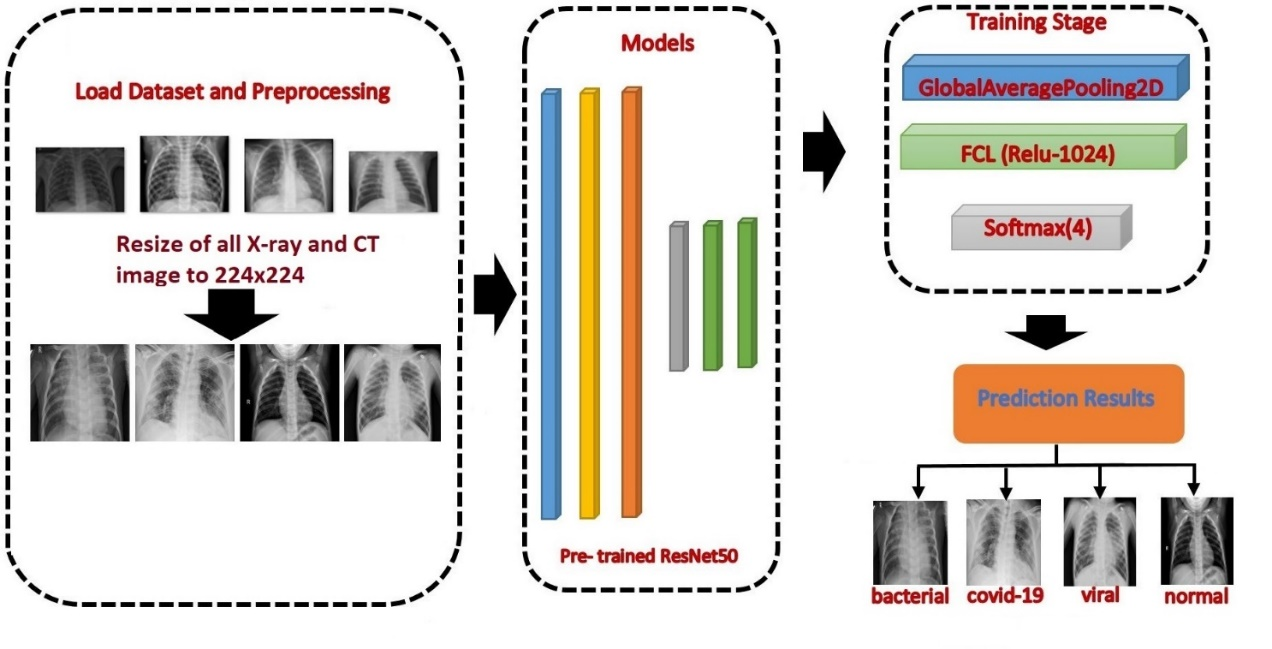}
\caption{Representation of the study.}
\label{figure1}
\end{figure}

\section{MATERIAL AND METHODS}

\subsection{Dataset}

Two different databases were used to obtain X-ray images. There are four different classification processes on X-Ray images. Of these, 219 images for COVID19 positive, 1341 images for normal and 1345 images for viral pneumonia were obtained from radiography-database images on Kaggle page \cite{Chowdhury2020}.  In addition to these data, 1350 X-ray images, which are pneumonia from bacteria, were also obtained from a different database on the Kaggle page \cite{Kaggle2020}. Normal, COVID19, Bacterial pneumonia and Viral pneumonia chest X-ray images are given in the Figure 2.

\begin{figure}[htbp]
\centering
\includegraphics[width=0.75\columnwidth]{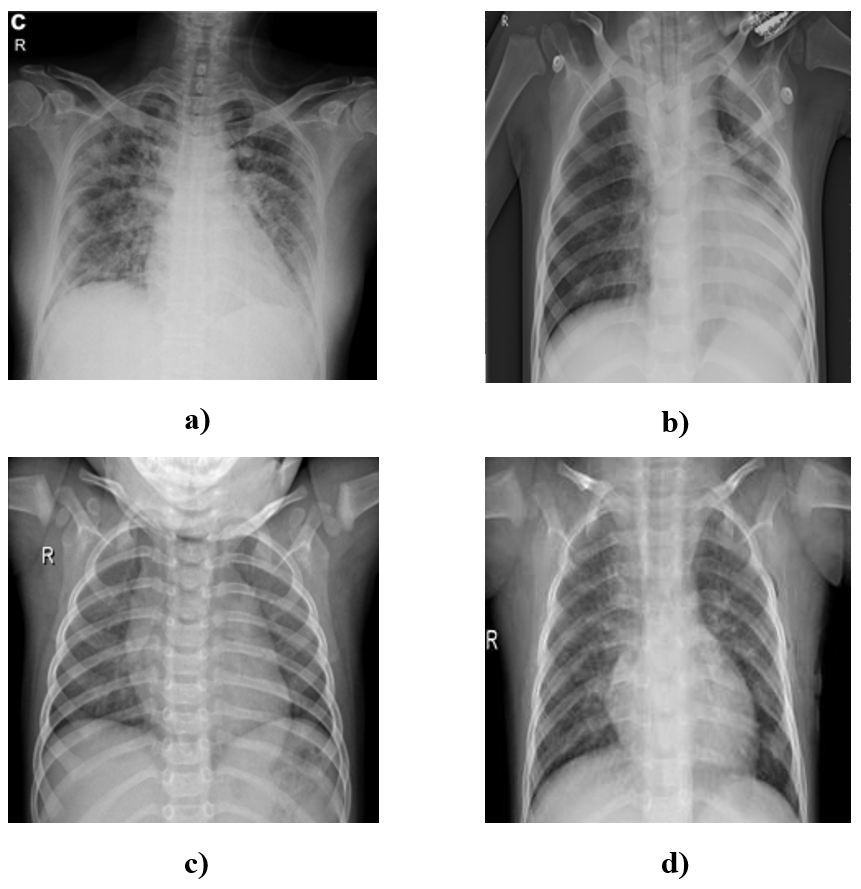}
\caption{X-Ray images of 4 different classes a) COVID19 b) Viral pneumonia c) Normal d) Bacterial pneumonia}
\label{figure1}
\end{figure}

\subsection{Convolutional Neural Networks}

CNN is a machine learning model. CNN differs from other neural network models by direct processing of data to produce results \cite{Acharya2018,NarinIRBM2020}  and is widely used in medical images because it can impress image processing for classification, detection and segmentation \cite{Han2018}. A standard CNN model usually includes convolution, pooling, and fully connected layers. The properties are obtained by the convolution layer with activation function. The pooling layer reduces the number of features and model parameters that reduce the number of calculations. In particular, the Relu activation function is used on the Fully Connected Layer, and the SoftMax activation function is added to the end of this layer and provides a categorical probability distribution output using CNN output.
The performances of deep learning models vary according to the hyper-parameters used. In order for the model to exhibit high performance, values such as data number, batch size value, learning rate, optimization algorithms, Epoch amount, activation functions, kernel size should be determined appropriately. In this study, the effect of bathsize parameters, which means how many data will be processed at the same time, will be examined.

\subsection{ResNet Model}

Kaiming He proposed Residual neural network (ResNet) in 2015 \cite{He2016}. The hypothesis behind ResNet is that deeper networks are more difficult to optimize. Because the deeper model should be able to perform as well as the shallower model by copying the learned parameters from the shallow model and adjusting the additional layers to identity mapping. More specifically, every residual block has two 3x3 convolutional layers. Periodically, the number of filters are doubled and spatial downsampling is operated.

\subsection{Metrics}

Predictive performances of classes are determined by four different measurements. The metrics are:

\begin{eqnarray}
  Accuracy&=&\frac{TP+TN}{TP+TN+FP+FN}\\
	Recall&=&\frac{TP}{TP+FN}\\
	Precision&=&\frac{TP}{TP+FP}\\
	F1-Score&=&\frac{2*Precision*Recall}{Precision+Recall}
\end{eqnarray}

where TP is the number of correctly classified positive subjects, FN is the number of misclassified positive subjects as negative, TN is the number of correctly classified negative subjects, and FP is the number of misclassified negative subjects as positive.




\section{RESULTS AND DISCUSSION}

Tensorflow-keras libraries supported by Python programming language were used for the training of the deep learning model. Original images have been resized and converted to 224x224. No other pre-processing was applied to the images. In updating the model weights, the ADAM algorithm, the Epoch number as 30 and the Learning rate as 0.0001 were determined. In addition to the direct transfer of the weights of the ResNet50 model, the model was trained with 3 layers added to the end of the model.
Training and test performance of six different batch size (3,10,20,30,40,50) values are given in Figure 3 and Figure 4.

\begin{figure}[htbp]
\centering
\includegraphics[width=1.0\columnwidth]{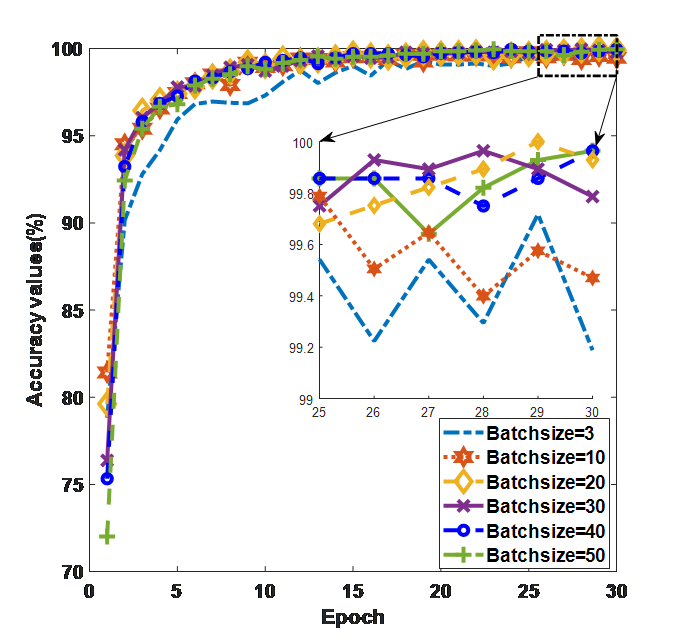}
\caption{The training performance of different BH parameters.}
\label{figure3}
\end{figure}

As can be seen from Figure 3, It is seen that the model exhibits an training performance over 99\% according to all BH parameters. In the zoomed graph, it is seen that low BH values get slightly lower.

\begin{figure}[htbp]
\centering
\includegraphics[width=1.0\columnwidth]{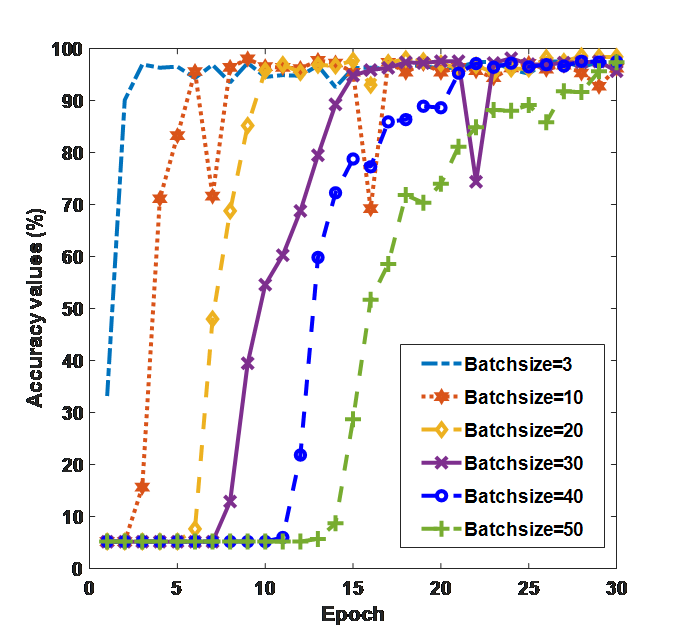}
\caption{Test performance of different BH parameters.}
\label{figure4}
\end{figure}

Considering Figure 4, it can be said that low GH values reach a stable result at low epoch (rapidly) values. As the BH value increases, the time to reach stability increases. This means that more epoch values may be needed. When we compare these results with the general performance values given in Table 1, the fact that the BH value is not too small and not too large makes the performance values higher. The highest COVID19 detection was 95.17\% for BH = 3 and the highest overall performance value and F1 score value were obtained for BH = 20.

\begin{table}[htbp]
  \centering
  \begin{tabular}{|c|c|c|c|c|}
\hline
Batch Size & Accuracy (\%) & Recall (\%) & Precision (\%) & F1-Measure (\%)\\
\hline
3	&97.73	&95.17	&94.41	&94.57\\
10	&95.29	&91.51	&88.64	&88.84\\
20	&97.97	&94.95	&95.98	&95.40\\
30	&97.82	&95.01	&95.70	&95.30\\
40	&97.93	&94.91	&95.96	&95.34\\
50	&95.66	&91.62	&92.83	&91.50\\
\hline
\end{tabular}
\caption{Obtained classification accuracies with different batch size.}
\label{tablo}
\end{table}

\end{document}